\documentclass[aps,prd,preprintnumbers,superscriptaddress,nofootinbib,floatfix,twocolumn,notitlepage,amsmath,amssymb]{revtex4-2}

\usepackage{graphicx}
\usepackage{dcolumn}
\usepackage{graphicx}
\usepackage{xcolor}
\usepackage{slashed}
\usepackage{xspace}
\usepackage[utf8]{inputenc}
\usepackage{float}
\usepackage{subfigure}
\usepackage{url}
\usepackage{color}
\usepackage[dvipsnames,table]{xcolor}
\usepackage{comment}
\usepackage{adjustbox}
\usepackage{booktabs} 
\usepackage{colortbl} 

\def\beq#1\eeq{\begin{align}#1\end{align}}

\newcommand{\Bbar}{\,\overline{\!B}}

\RequirePackage{xspace}
\def\Bbar    {\kern 0.18em\overline{\kern -0.18em B}{}\xspace}

\definecolor{BlueViolet}{rgb}{0.2, 0.00, 0.7}
\definecolor{Blue}{rgb}{0.15, 0.00, 0.9}
\definecolor{halayaube}{rgb}{0.4, 0.22, 0.33}
\definecolor{sanddune}{rgb}{0.59, 0.44, 0.09}

\usepackage[colorlinks=true,linkcolor=Blue,citecolor=Blue,urlcolor=BlueViolet]{hyperref}
\usepackage[hyphenbreaks]{breakurl} 

\begin{document} 

\title{\bf Role of heavy neutral lepton in lepton number violating $B$ meson decays}
\author{Dhiren Panda}
\email{pandadhiren530@gmail.com}
\author{Manas Kumar Mohapatra}
\email{manasmohapatra12@gmail.com}
\author{Rukmani Mohanta}
\email{rmsp@uohyd.ac.in}
\affiliation{School of Physics,  University of Hyderabad, Hyderabad-500046,  India
}


\begin{abstract}
\noindent
We study the phenomenology of heavy neutral leptons (HNLs) in $B$-meson decays as probes of physics beyond the Standard Model. Focusing on the leptonic channels $B \to \mu N$ and $B \to \tau N$, we constrain the allowed regions in the $M_N$--$|U_{\ell N}|^2$ plane using current experimental data. Using these constraints, we investigate lepton-number violating ($\Delta L=2$) processes mediated by on-shell HNLs, including $B_{(c)}^- \to \pi^+ \mu^- \mu^-$ and $B_c^- \to J/\psi\, \pi^+ \mu^- \mu^-$. For benchmark values $|U_{\mu N}|^2 = 10^{-6}$ and $M_N = 2$-- $3\,\mathrm{GeV}$, the predicted branching ratios lie in the range $\mathcal{O}(10^{-13})$--$\mathcal{O}(10^{-8})$. Among the channels, $B_c^- \to \pi^+ \mu^- \mu^-$ shows the largest enhancement, while $B_c^- \to J/\psi\, \pi^+ \mu^- \mu^-$ is strongly suppressed. These results indicate a clear channel dependence, with $B_c$ modes providing enhanced sensitivity to HNL effects and offering promising avenues for future searches of lepton number violation.

\end{abstract}
\maketitle
\section{Introduction}
The discovery of the Higgs boson at Large Hadron Collider (LHC)~\cite{ATLAS:2012yve, CMS:2012qbp} completes the particle content of the Standard Model (SM) with three active neutrino species as massless. However, the observation of neutrino oscillations~\cite{Super-Kamiokande:1998kpq,SNO:2002tuh,KamLAND:2002uet, DayaBay:2012fng} demonstrates that neutrinos have nonzero masses, providing clear evidence of physics beyond the SM. Nevertheless, several fundamental properties of neutrinos remain unknown, including whether they are Dirac or Majorana particles, their absolute mass scale, the ordering of their masses, and the extent of CP violation in the lepton sector etc. It is also unclear whether for generating neutrino masses require additional particles beyond the SM. \\

To address these open questions, various extensions of the SM have been proposed. In this context, heavy neutrinos are well-motivated candidates for generating neutrino masses. In particular, the right-handed neutrinos with Majorana masses give rise to a spectrum of three light active neutrinos accompanied by heavy Majorana states~\cite{Asaka:2016rwd}. These heavy states are typically expected to lie above the electroweak scale, potentially near the grand unification scale ($\gtrsim 10^{6}$~GeV), where they can account for the observed baryon asymmetry of the Universe through leptogenesis~\cite{Fukugita:1986hr,Asaka:2005pn}.
On the other hand, neutrinos with masses below the $W$ boson mass are also of considerable interest. In this regime, the seesaw mechanism remains viable provided that the corresponding neutrino Yukawa couplings are sufficiently suppressed \cite{Asaka:2016rwd}. 
Moreover, heavy neutrinos with masses around 100 MeV have been investigated in the context of supernova dynamics \cite{Fuller:2008erj}, while those at the keV scale constitute well-motivated dark matter candidates \cite{Dodelson:1993je}. Interestingly, GeV-scale HNLs can play an important role in low-energy flavor observables and collider phenomenology.

A particularly striking consequence of Majorana neutrinos is the violation of lepton number by two units ($\Delta L = 2$). Such lepton-number violating (LNV) processes provide a powerful probe of the Majorana nature of neutrinos and have been extensively studied in various meson and lepton decays. In particular, three-body decays of charged mesons, such as $(K^-, D^-_{(s)}, B^-, B^-_c) \to M^+ \ell^- \ell'^-$, as well as $\tau$-lepton decays, have been widely investigated as promising channels to search for LNV signatures~\cite{Ali:2001gsa,Atre:2005eb,Ivanov:2004ch,Helo:2010cw,Cvetic:2010rw,Zhang:2010um,Bao:2012vq,Wang:2014lda,Atre:2009rg,Gribanov:2001vv}. Complementary studies have also considered the four-body decays of heavy mesons and $\tau$ leptons as additional probes of such effects~\cite{Quintero:2011yh,Castro:2013jsn,LopezCastro:2012udb, Dong:2013raa,Yuan:2013yba,LopezCastro:2012rbs,LopezCastro:2012udb,Dib:2011hc}. Experimental searches for these processes have been carried out by several Collaborations, including Belle, LHCb, BABAR, and E791~\cite{BELLE:2011bej,LHCb:2014osd,LHCb:2011yaj,BaBar:2013swg,E791:2000jkj}, but no conclusive evidence for LNV signal has been observed so far. 

Despite these extensive studies, purely leptonic decays of $B$ mesons involving heavy neutral leptons remain less explored, especially as probes of mixing structure and as complementary constraints on HNL parameter space. These processes are theoretically clean and sensitive to new physics, with precisely predicted SM branching ratios that can show sizable deviations in the presence of heavy neutrinos. In this work, we investigate the role of heavy neutral leptons in purely leptonic $B$ meson decays. In particular, we focus on the decay channels $B^\pm \to \ell^\pm N$, where $N$ denotes a heavy neutral lepton that mixes with the SM neutrinos. We further explore the implications of these processes on the lepton-number violating decays mediated by on-shell HNLs, such as $B^- \to \pi^+ \mu^- \mu^-$, $B^-_c \to \pi^+ \mu^- \mu^-$, and $B^-_c \to J/\psi  \pi^+ \mu^- \mu^-$. The LNV $B$ decays have been extensively studied, particularly at LHCb. Searches for $B^- \to \pi^+ \mu^- \mu^-$ have reported no signal, with upper limits on the branching fraction at $\mathcal{O}(10^{-9})$ for heavy neutrino masses in the range $250~\text{MeV} \lesssim M_N \lesssim 5~\text{GeV}$~\cite{LHCb:2014osd}. Similarly, the branching fractions might be accessible at the order $\leq (10^{-7} - 10^{-8})$ at LHCb through on-shell heavy neutrino contributions~\cite{Milanes:2016rzr}. 
These results demonstrate that such channels provide promising and experimentally accessible probes of heavy neutral leptons in the MeV–GeV mass range.
Overall, these channels offer complementary and powerful avenues to probe the Majorana nature of neutrinos and to test the existence of heavy neutral leptons at current and future high-energy experiments.
Recently, the LHCb experiment has reported upper bounds on the $B^- \to D^{(*)+}\mu^-\mu^-$ decay channels at the level of $\mathcal{O}(10^{-8})$~\cite{LHCb:2026evs}. In the extension of the Standard Model with heavy Majorana neutrinos, these modes are strongly suppressed, arising from annihilation-dominated contributions and yielding branching ratios well below current experimental sensitivity. We therefore do not include this channel in our analysis and focus instead on $B_{(c)}$-meson decays, which provide experimentally favorable access to resonant heavy neutrino effects.

This paper is organized as follows. In Section \ref{BtoDlN}, we present the theoretical framework involving a light sterile neutrino. In Section \ref{sec:num_disc1}, we explicitly review the lepton-number violating (LNV) three- and four-body decay channels. In Section \ref{sec:num_disc}, we present the details of the allowed parameter space consistent with current experimental constraints, followed by a numerical analysis and a discussion of the feasibility of the LNV decay and the corresponding detector sensitivity. Finally, in Section \ref{Conclu_1}, we summarize our conclusions.

\section{$B$ meosn decays with heavy neutrinos}
In this section, we explore the semileptonic $B$ meson decays in the presence of heavy neutral leptons. We first consider purely leptonic decays $B \to \ell N$, followed by lepton-number violating processes in the presence of on-shell Majorana neutrinos.

In general, the neutrino flavor eigenstates can be expressed as linear combinations of light and heavy mass eigenstates. In the presence of a single heavy neutral lepton $N$, the flavor eigenstates $\nu_\ell$ ($\ell = e, \mu, \tau$) are given by
\begin{equation}
\nu_\ell = \sum_{i=1}^{3} U_{\ell \nu_i}\,\nu_i + U_{\ell N}\,N,
\end{equation}
where $\nu_i$ denote the light mass eigenstates and $U_{\ell \nu _i}$ are the elements of the standard $3 \times 3$ Pontecorvo–Maki–Nakagawa–Sakata (PMNS) matrix. The parameter $U_{\ell N}$ describes the mixing between the heavy state and the Standard Model neutrinos. In this framework, the full leptonic mixing matrix is a $4 \times 4$ unitary matrix, which leads to an effective non-unitarity of the $3 \times 3$ PMNS sub-matrix:
\begin{equation}
\sum_{k=1}^{3} \left| U_{\ell \nu_ k} \right|^2
= 1 - \left| U_{\ell N} \right|^2.
\label{eq:unitarity_condition}
\end{equation}
This framework provides the basis for our analysis.

\subsection{Leptonic $B$ Decays with Heavy Neutral Leptons}
\label{BtoDlN}
For the purely leptonic decay $B^+ \to \ell^+ \nu_\ell$ ($\ell = e, \mu, \tau$), the SM predictions are helicity suppressed, and in the limit of a massless neutrino, the decay width is given by~\cite{Cvetic:2017gkt}
\begin{equation}
\Gamma_{\rm SM}(B^+ \to \ell^+ \nu_{\ell}) =
\frac{G_F^2 f_B^2 |V_{ub}|^2 M_B^3}{8\pi}\,
y_{\ell} (1 - y_{\ell})^2 ,
\end{equation}
where $y_{\ell} = M_{\ell}^2/M_B^2$. In scenarios with a heavy neutral lepton $N$, the leptonic decay channel $B^+ \to \ell^+ N$ becomes kinematically accessible and contributes to the inclusive leptonic signature. The corresponding decay width factorizes into the heavy–light mixing parameter with a purely kinematic term, and is given as~\cite{Cvetic:2017gkt}
\begin{equation}
\Gamma(B^+ \to \ell^+ N)
=
|U_{\ell N}|^2 \,
\overline{\Gamma}(B^+ \to \ell^+ N), \label{eq:4}
\end{equation}
where $U_{\ell N}$ denotes the active–sterile neutrino mixing element.
The function $\overline{\Gamma}$ encodes the phase-space effects associated with the finite mass of the heavy neutrino and is expressed as ~\cite{Cvetic:2017gkt}
\begin{align}
&\overline{\Gamma}(B^+ \to \ell^+ N)
= \frac{G_F^2 f_B^2 |V_{ub}|^2 M_B^3}{8\pi}\,
\lambda^{1/2}(1,y_N,y_\ell) \nonumber \\
&~~~~~~~\quad \times
\Big[(1 - y_N) y_N + y_\ell (1 + 2 y_N - y_\ell)\Big], \label{eq:5}
\end{align}
with $y_N = M_N^2/M_B^2$ and
\[
\lambda(x,y,z) = x^2 + y^2 + z^2 - 2xy - 2yz - 2zx .
\]

In the limit $y_N \to 0$, the above expression reduces to the Standard Model result, providing a consistency check. Experimentally, the inclusive leptonic signal $B^+ \to \ell^+ + \text{``missing momentum"}$ receives contributions from both light neutrinos and possible heavy neutrino states. Under the assumption of unitarity of extended mixing matrix, the total rate can be written as
\begin{align}
&\Gamma (B^+ \to \ell^+ + \text{mm})
= \Gamma_{\rm SM}(B^+ \to \ell^+ \nu) \nonumber \\
&~~~~~~~~~+ |U_{\ell N}|^2 \Big[
\overline{\Gamma}(B^+ \to \ell^+ N) 
 - \overline{\Gamma}(B^+ \to \ell^+ \nu)
\Big],
\end{align}
with mm = missing momentum.
Without the unitarity condition, the result corresponds to an incoherent sum of light- and heavy-neutrino contributions:

\begin{align}
\Gamma(B^+ \to \ell^+ + \text{mm})
&= \Gamma(B^+ \to \ell^+ \nu) \nonumber\\
&+ |U_{\ell N}|^2 \, \overline{\Gamma}(B^+ \to \ell^{+} N).
\end{align}

\subsection{Lepton Number Violating $B$ Meson Decays }
\label{sec:num_disc1}
 \begin{figure}
	\centering
	\includegraphics[width=0.5\textwidth]{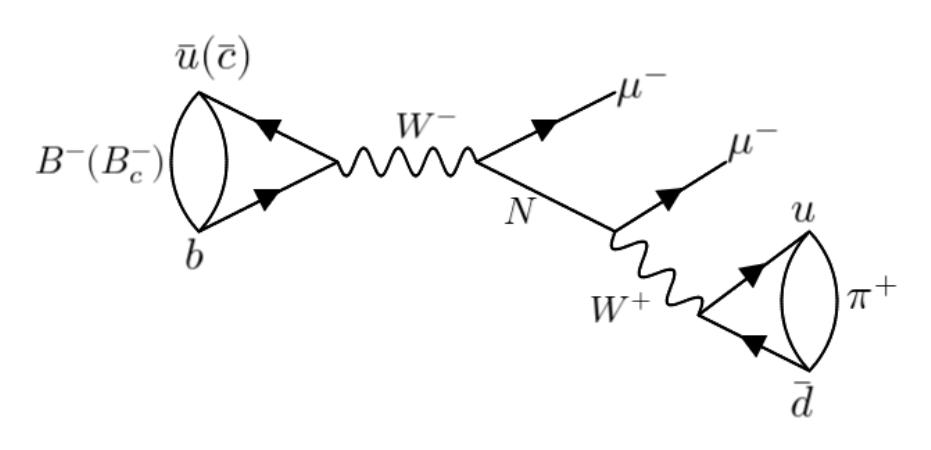}
        \includegraphics[width=0.45\textwidth]{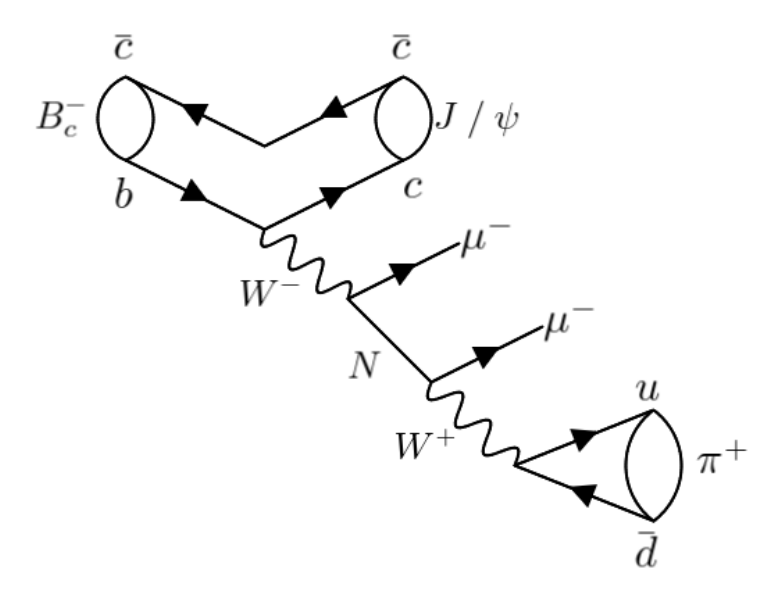}
	\caption{\small{Feynman diagrams for LNV meson decays with possible resonant enhancement.}}
	\label{feynman_LNV}
\end{figure}
The lepton number violating $B$ meson decays can arise via the exchange of an on-shell Majorana neutrino within the kinematically accessible region, leading to same-sign dilepton final states $\ell_1^- \ell_2^-$ ($\ell_1,\ell_2 = e,\mu,\tau$). We consider a scenario with a single heavy neutrino $N$ produced on-shell and dominates the decay amplitude. In this sub-section, we focus on the dimuon channel and study the three-body decays $B_c^- \to \pi^+ \mu^- \mu^-$, $B^- \to \pi^+ \mu^- \mu^-$, as well as the four-body decay $B_c^- \to J/\psi\, \pi^+ \mu^- \mu^-$. The corresponding Feynman diagrams are shown in Fig.~\ref{feynman_LNV}. We begin with the three-body decays, followed by the four-body processes as follows. 
\subsubsection{ $B_{(c)}^- \to \pi^+\mu^-\mu^-$ } \label{Bc3body}
The decays $B_c^- \to \pi^+ \mu^- \mu^-$ and $B^- \to \pi^+ \mu^- \mu^-$ proceed via resonant production of an intermediate Majorana neutrino $N$, factorizing into the sequential processes
\begin{equation}
B_{(c)}^- \to \mu^- N, \qquad
N \to \mu^- \pi^+ .
\end{equation}
The decay rate is resonantly enhanced when $N$ is produced on-shell, which requires
\begin{equation}
m_\pi + m_\mu < M_N < m_{B_{(c)}} - m_\mu .
\end{equation}
Numerically, this corresponds to
\begin{align}
0.25~\mathrm{GeV} \le M_N \le 6.16~\mathrm{GeV}
\quad \text{for } B_c^- \to \pi^+\mu^-\mu^- , \nonumber\\
0.25~\mathrm{GeV} \le M_N \le 5.17~\mathrm{GeV}
\quad \text{for } B^- \to \pi^+\mu^-\mu^- .
\end{align}

Within the narrow-width approximation ($\Gamma_N \ll M_N$)~\cite{Atre:2009rg}, the total decay rate factorizes into the product of the two sub-processes divided by the total width of the intermediate state as,
\begin{equation}
\Gamma(B_{(c)}^- \to \pi^+\mu^-\mu^-)
=
\frac{\Gamma(B_{(c)}^- \to \mu^- N)\,\times
\Gamma(N \to \pi^+\mu^-)}{\Gamma_N}. 
\label{commonfactor}
\end{equation}
Here the expression of the total decay width $\Gamma _N$ is given in Appendix \ref{Total decay width of N}.
The corresponding partial width for $B^- \to \mu^-N$ is given in Eqs. (\ref{eq:4}) and (\ref{eq:5}), while for $N \to \pi^+ \mu^-$, it takes the form
\begin{eqnarray}
\label{decaywidths}
%
\Gamma(N \to \pi^+\mu^-)
&=&
\frac{G_F^2}{16\pi}
f_\pi^2 M_N^3
|V_{ud}|^2 |U_{\mu N}|^2 \lambda^{1/2}(1,x_\pi,x_\mu)\nonumber\\
& \times & \left[1-x_\pi-2 x_\mu -x_\mu(x_\pi -x_\mu)\right] ,
\end{eqnarray}
where $x_{\pi/\mu}={m_{\pi/\mu}^2}/{M_N^2}$, $V_{qb}=V_{cb}$ for $B_c^-$ and $V_{qb}=V_{ub}$ for $B^-$ decays, and $f_{B_{(c)}}$ and $m_{B_{(c)}}$ denote the decay constant and mass of the parent mesons, respectively.
\subsubsection{$B_c^- \to J/\psi \pi^+\mu^-\mu^-$}   \label{Bc4body}
This channel probes a kinematically accessible Majorana neutrino with mass in the range $0.25~\mathrm{GeV} \le M_N \le 3.18~\mathrm{GeV}$. Unlike the three-body channel, the four-body decay exhibits distinct kinematic and dynamical features, thereby providing an additional LNV probe of the heavy neutrino sector. The diagram contributing to the four-body decay $B_c^- \to J/\psi\, \pi^+ \mu^- \mu^-$ is shown in the bottom panel of Fig.~\ref{feynman_LNV}.

Following the same narrow-width approximation arguments as in the case of three-body decays, the Majorana neutrino is treated as an on-shell intermediate state produced in the semileptonic decay $B_c^- \to J/\psi\, \mu^- N$, followed by the subsequent decay $N \to \mu^- \pi^+$. Within this on-shell factorization framework, the decay width can be written as
\begin{equation}
\begin{aligned}
\Gamma(B_c^- \to J/\psi\, \pi^+ \mu^- \mu^-)
&= \frac{\Gamma(B_c^- \to J/\psi\, \mu^- N)}{\Gamma_N} \\
&\times \Gamma(N \to \mu^- \pi^+)\;,
\end{aligned}
\label{BcJpsiDecay}
\end{equation}
where $\Gamma(N \to \mu^- \pi^+)$ is given in Eq.~\eqref{decaywidths}. In the limit of a negligible charged-lepton mass, the decay width of the sub-process $B_c^- \to J/\psi\, \mu^- N$ can be expressed as
\begin{equation}
\begin{split}
\Gamma(B_c^- \to J/\psi\, \mu^- N)
= \frac{G_F^2}{32 (2\pi)^3 m_{B_c}^3}
|V_{cb}|^2 |U_{\mu N}|^2 \\
\times \int_{M_N^2}^{(m_{B_c} - m_{J/\psi})^2} dt \;
\mathcal{A}(t).
\end{split}
\label{BctopimuN}
\end{equation}
where $\mathcal{A}(t)$ encodes the underlying hadronic dynamics as a function of the momentum transfer $t = (p_{B_c}-p_{J/\psi})^2$, as detailed in Eq.~\eqref{At}.
\section{Numerical Analysis}
\label{sec:num_disc}
In this section, we present the numerical analysis of the relevant decay rates, focusing on their dependence on the heavy neutrino mass and mixing parameters, and explore the resulting phenomenological implications. We begin with the leptonic decay channels $B \to \mu N$ and $B \to \tau N$, which play a central role in the analysis, and confront our predictions with current experimental constraints.
\subsection{Constraints on the $|U_{\ell N}|^2$–$M_N$ plane from leptonic $B$ decays }
\label{sub:num_RD_RDst}
\begin{figure}[t]
    \centering
    \includegraphics[width=0.46\textwidth,height=7cm]{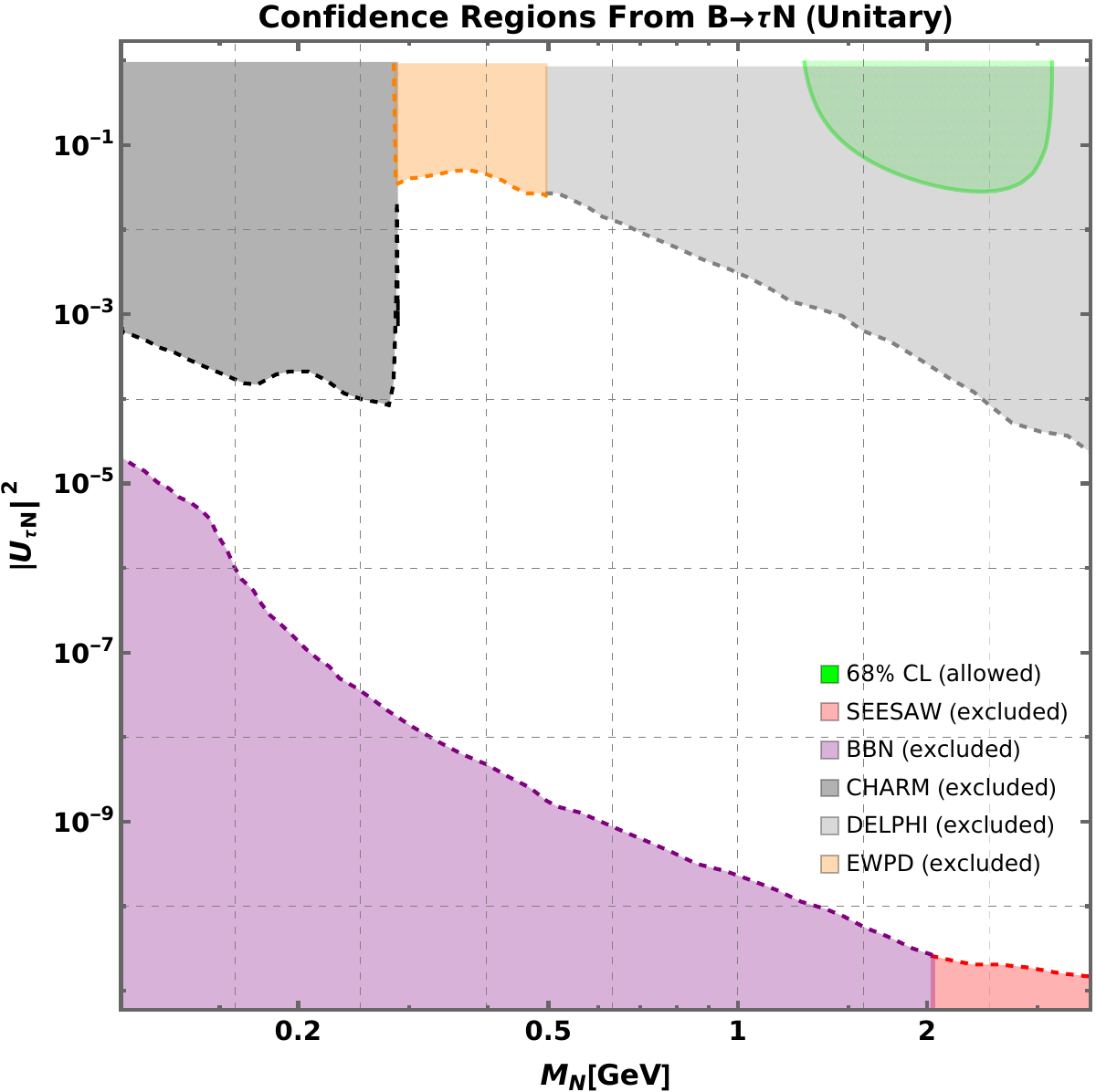}\\[0.6cm]
    \includegraphics[width=0.46\textwidth,height=7cm]{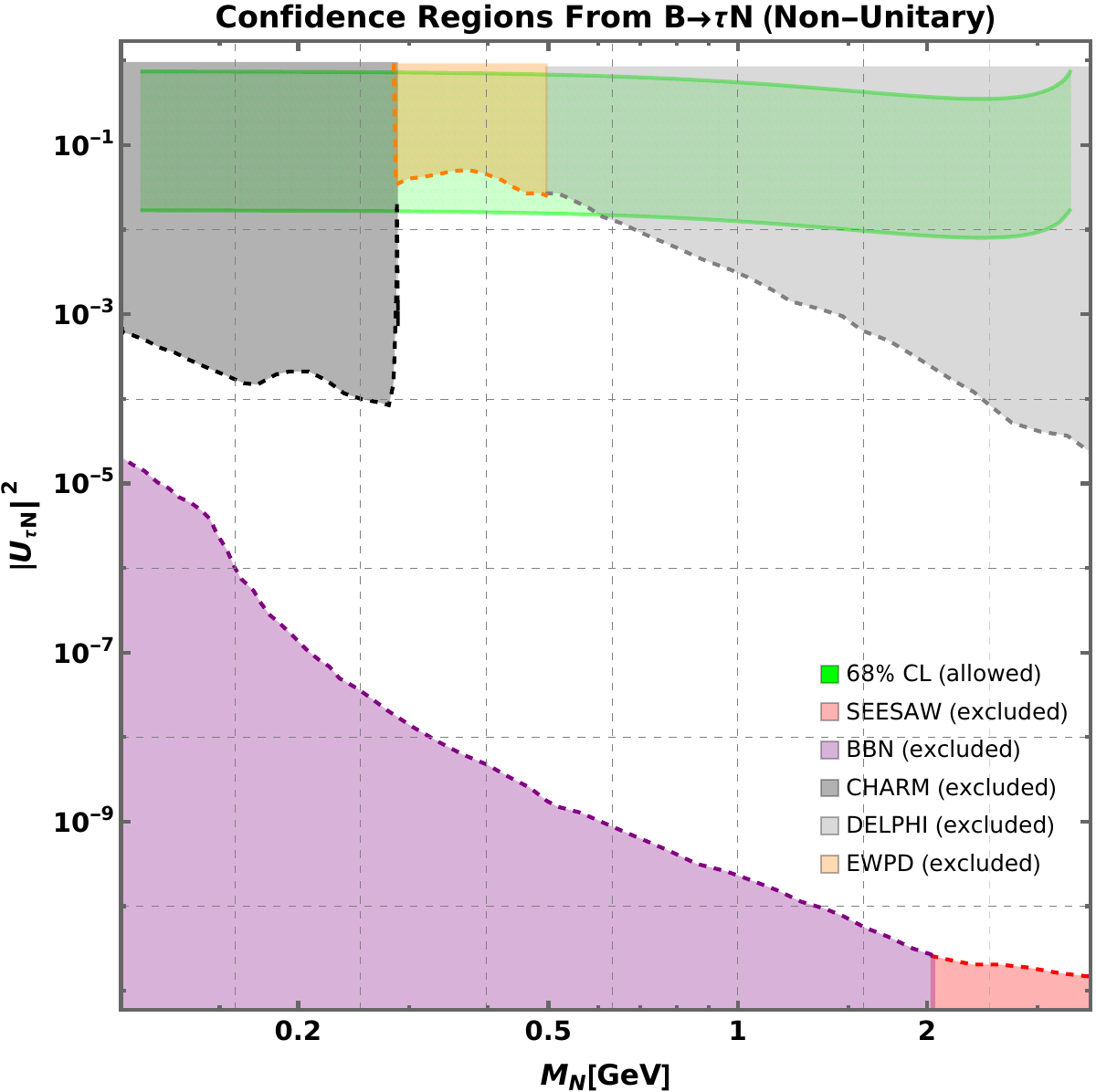}

    \caption{Confidence regions for the HNL mixing parameter $|U_{\tau N}|^2$ as a function of $M_N$ in $B \to \tau N$ decays, under unitary (top) and non-unitary (bottom) assumptions. The green shaded area denotes the 68\% CL allowed region, with excluded regions from SEESAW \cite{deGouvea:2009fp,deGouvea:2005er,Cirelli:2004cz}, BBN \cite{Gorbunov:2007ak,Ruchayskiy:2012si}, CHARM \cite{Orloff:2002de}, DELPHI \cite{DELPHI:1997qjz}, and EWPD \cite{delAguila:2008pw,Akhmedov:2013hec,deBlas:2013gla} constraints.}
    \label{fig:B_tn_decays}
\end{figure}

\begin{figure}[htbp]
    \centering
    \includegraphics[width=0.46\textwidth,height=7cm]{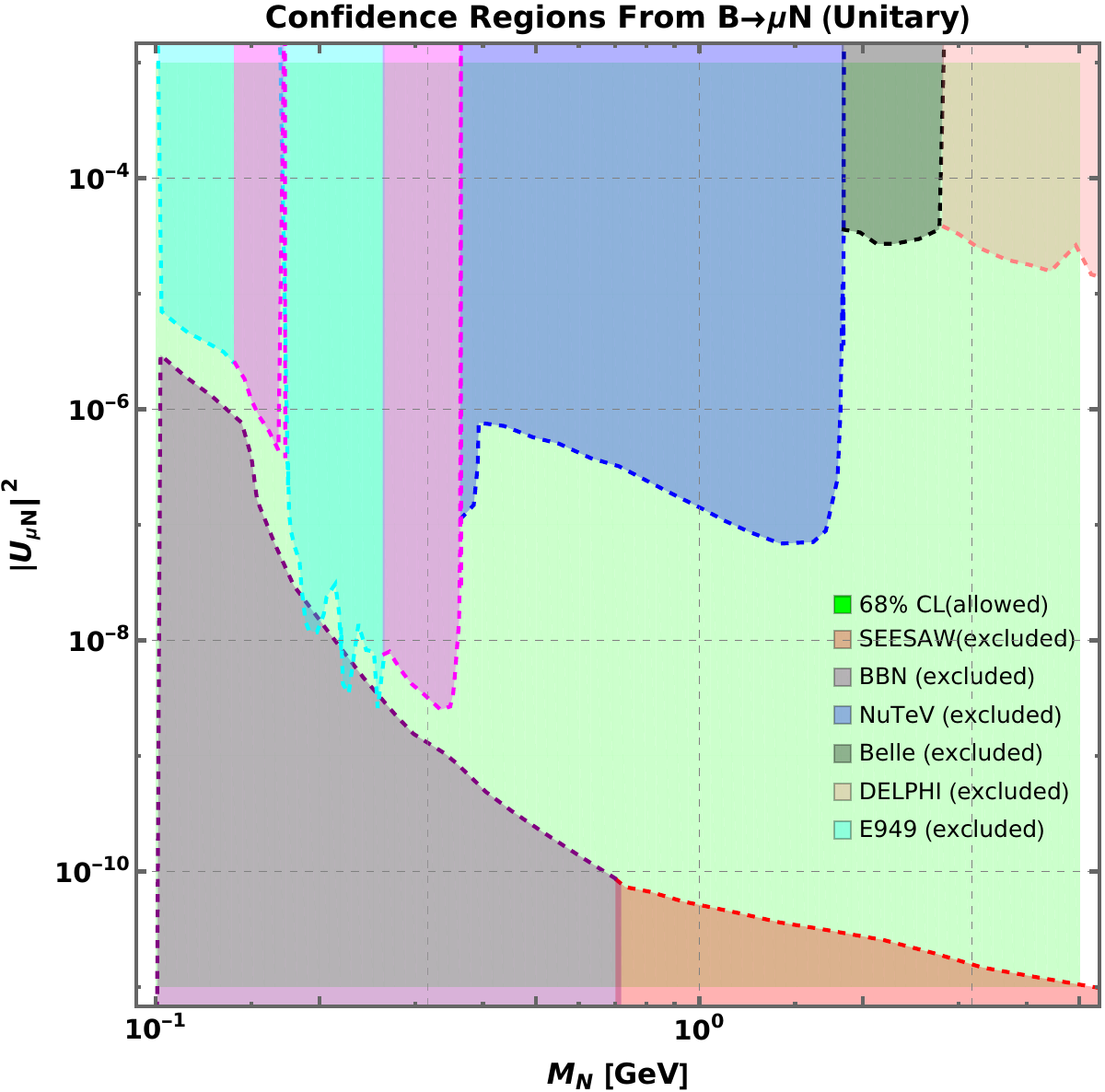} \\[0.6cm]
     \includegraphics[width=0.46\textwidth,height=7cm]{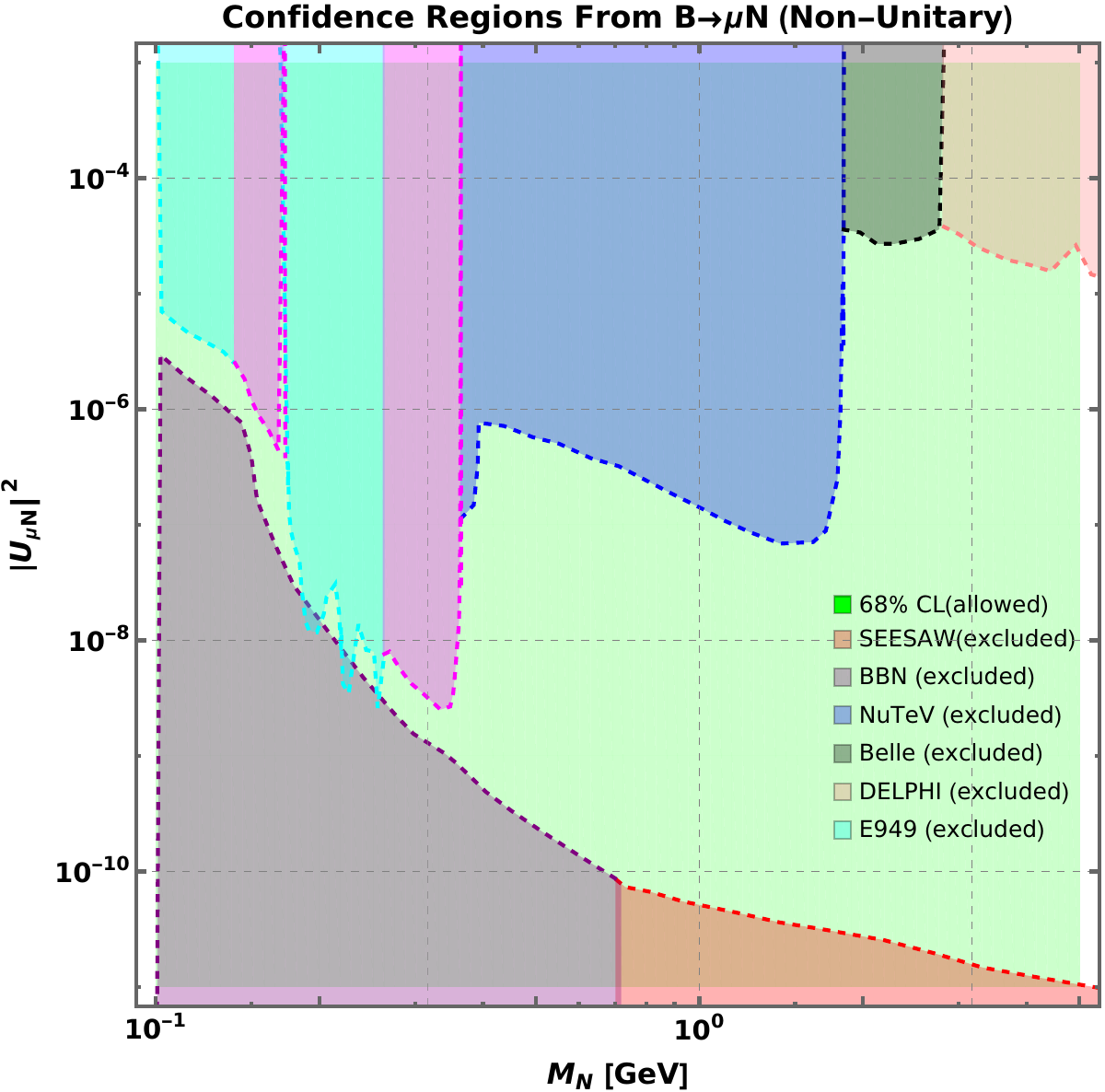}%
  \caption{Confidence regions for $|U_{\mu N}|^2$ versus $M_N$ in $B \to \mu N$ decays under unitary (top) and non-unitary (bottom) assumptions. The green region denotes the 68\% CL allowed area, with exclusions from SEESAW \cite{deGouvea:2009fp,deGouvea:2005er,Cirelli:2004cz}, BBN \cite{Gorbunov:2007ak,Ruchayskiy:2012si}, NuTeV \cite{NuTeV:1999kej}, Belle \cite{Belle:2013ytx}, DELPHI \cite{DELPHI:1997qjz}, and E949 \cite{E949:2014gsn}.}
  \label{fig:B_mun_decays}
\end{figure}

In this sub-section, we present a detailed analysis of the allowed parameter space implied by our results. Using the purely leptonic decay $B \to \ell \nu~(\ell=\mu,\tau) $ process, we constrain the active-sterile neutrino mixing parameter $|U_{\ell N}|^2$ as a function of heavy neutrino mass $M_N$. The experimental branching fraction limits are given in the Table \ref{tb1}.
\begin{table}[h]
\caption{Branching fractions of $B^+ \to  ~\mu^+ \nu, ~\tau^+ \nu$ in units of $10^{-6}$~\cite{Belle-II:2025ruy,Belle-II:2026flt,Cvetic:2017gkt}.}
\label{tb1}
\smallskip
\begin{tabular}{c  c c}
\hline \hline
$B^+ \to$  &    $\mu^+ \nu$  &  $\tau^+ \nu$  \\
\hline
BABAR  &    ~~$< 1.0$~~  &  ~~$179 \pm 48$ \\
Belle I and II &    ~~$< 0.67$~~  &  ~~$124 \pm 41 \pm 11$ \\
Experimental average  &    $< 1.0$  &  $106 \pm 19$  \\
\hline \hline
\end{tabular}
\end{table}
The decay modes $B \to \tau \nu$ and $B \to \mu \nu$ constrain the active--sterile mixing parameters $|U_{\tau N}|^2$ and $|U_{\mu N}|^2$, respectively, thereby probing different leptonic sectors.
The allowed regions in the $|U_{\ell N}|^2$--$M_N$ plane are determined from the branching ratios, including heavy neutrino contributions, and by requiring consistency with current experimental limits. These regions, together with existing exclusion bounds from other experiments, are shown in Fig.~\ref{fig:B_tn_decays} and Fig.~\ref{fig:B_mun_decays}. The comparison highlights the complementary sensitivity of purely leptonic $B$-meson decays in probing the heavy Majorana neutrino parameter space.
These processes depend on the heavy--light mixing parameter $U_{\mu N}$, which governs the coupling of the heavy Majorana neutrino to the muon sector and thus determines $|U_{\mu N}|^2$.
For completeness, we also scrutinize the decay mode $B \to \tau \nu$ which would provide complementary constraints on $|U_{\tau N}|^2$. 

In our analysis, we further consider two scenarios: (i) a unitary mixing case, where the extended lepton mixing matrix satisfies unitarity and deviations from the standard three flavour PMNS structure are consistently incorporated, and (ii) a non-unitary case, where no unitarity constraint is imposed and the heavy neutrino contribution modifies decay amplitudes independently.
This classification allows us to systematically study the impact of different mixing assumptions on the allowed parameter space and the decay rates of LNV $B$-meson processes.

For the $B \to \tau \nu$ channel, Fig.~\ref{fig:B_tn_decays} shows the allowed region in the $|U_{\tau N}|^2$--$M_N$ plane, shown in green, at 68\% C.L. Most of the parameter space is already excluded by existing bounds, leaving only a small residual region associated with non-unitary mixing. However, this surviving region typically leads to enhanced LNV branching ratios that exceed current experimental limits, rendering it phenomenologically disfavoured. Consequently, the $\tau$ channel does not provide additional viable constraints for our analysis of same-sign dimuon LNV $B$ decays.

In contrast, the $B \to \mu \nu$ channel, shown in Fig.~\ref{fig:B_mun_decays}, exhibits nearly identical allowed regions in both unitary and non-unitary scenarios, as the decay is strongly helicity suppressed and thus only weakly sensitive to non-unitarity effects. The colored regions indicate exclusions from various experiments, while the green region represents the currently allowed parameter space. 

It should be noted that the constraints obtained from $B \to \tau \nu$ and $B \to \mu \nu$ processes  provide bounds on the mixing parameters  $|U_{\tau N}|^2$ and $|U_{\mu N}|^2$, respectively. However, these two constraints lead to significantly different allowed regions, implying the violation of universal mixing scenario, i.e.,  $|U_{\tau N}|^2 \simeq |U_{\mu N}|^2$. Therefore, instead of enforcing universal mixing, we treat the muon sector independently, which is more relevant for the $M_1 \to M_2 \mu^-\mu^-$ motivated framework and provides a consistent description of the $B$ meson observables studied here.
\subsection{Impact on $\Delta L = 2$ $B$-meson decays}
\begin{figure*}[t]
    \centering
    \includegraphics[width=0.8\textwidth,height=9cm]{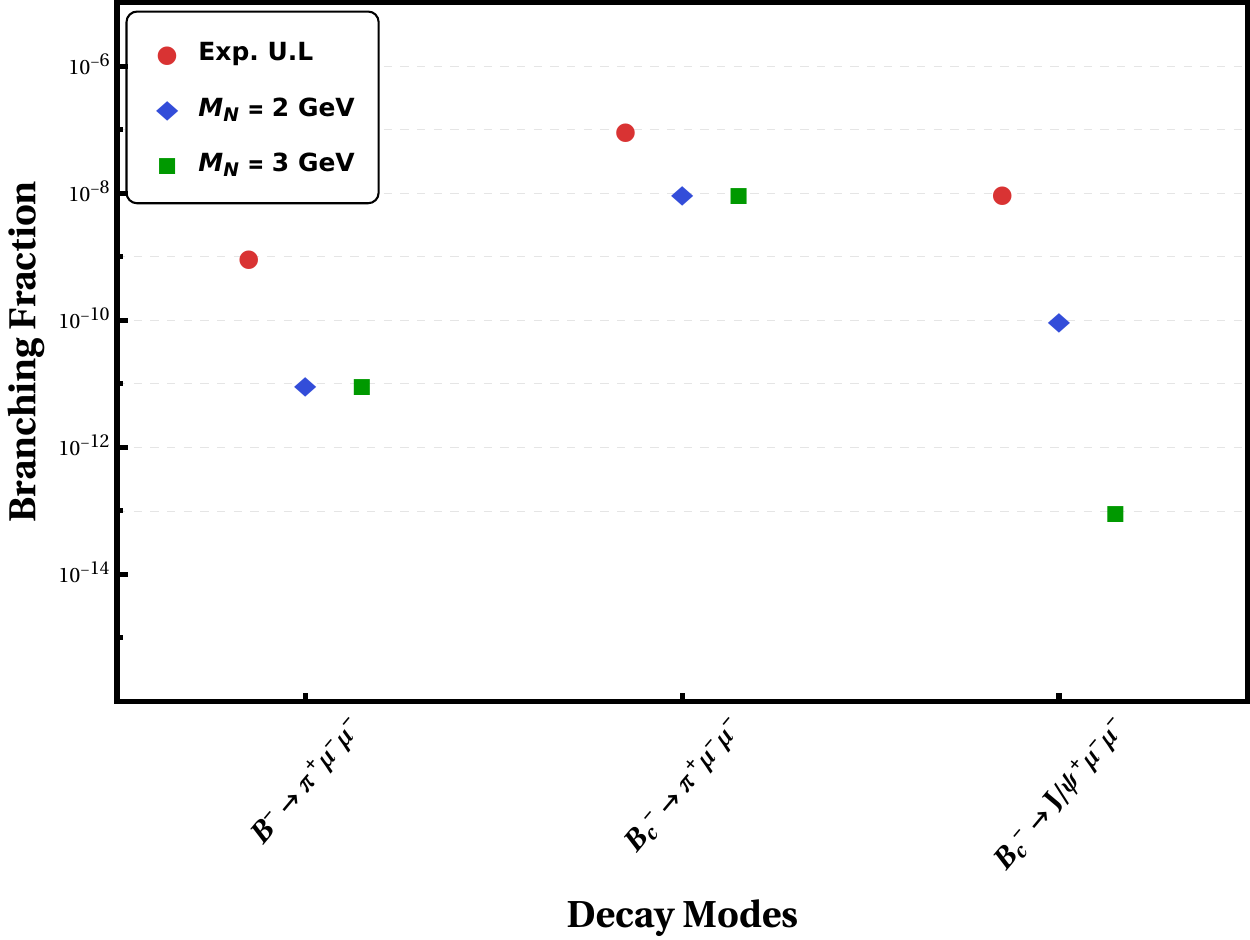}
  \caption{Comparison of limits and predictions for the considered decay modes. For $B^- \to \pi^+ \mu^- \mu^-$, the orange point denotes the experimental upper limit, while the blue and green points show our predictions for $M_N = 2$ and $3~\mathrm{GeV}$. For the other two modes, the orange points are taken from bounds in the literature\cite{Milanes:2016rzr}, with blue and green again representing predictions for $M_N = 2$ and $3~\mathrm{GeV}$.}
    \label{fig:LNV}
\end{figure*}

Using the $B \to \mu \nu$ channel, we select benchmark points $|U_{\mu N}|^{2}=10^{-6}$ and $M_{N}=2$ and 3 GeV for the numerical analysis. The resulting predictions for LNV $B$ meson decays are summarized in Table~\ref{tab:decay_limits} and shown in Fig.~\ref{fig:LNV}.
\begin{table}[h!]
\centering
\begin{tabular}{|c|c|c|}
\hline
Decay Mode ($|U_{\mu N}|^2 = 10^{-6}$) & $M_N = 2$ GeV & $M_N = 3$ GeV \\
\hline
$B^- \to \pi^+ \mu^- \mu^-$ 
& $\mathcal{O}(10^{-11})$ 
& $\mathcal{O}(10^{-11})$ \\
\hline
$B_c^- \to \pi^+\mu^-\mu^-$
& $\mathcal{O}(10^{-8})$ 
& $\mathcal{O}(10^{-8})$ \\
\hline
$B_c^- \to J/\psi \pi^+\mu^-\mu^-$
& $\mathcal{O}(10^{-10})$ 
& $\mathcal{O}(10^{-13})$ \\
\hline
\end{tabular}
\caption{Predicted upper limits on branching ratios for different decay modes 
for a heavy--light neutrino mixing $|U_{\mu N}|^2 = 10^{-6}$. 
Results are presented for heavy neutrino masses $M_N = 2$ GeV and $3$ GeV.}
\label{tab:decay_limits}
\end{table}

The predicted branching ratios lie in the range $\mathcal{O}(10^{-13})$--$\mathcal{O}(10^{-8})$, indicating strong suppression, as expected for rare LNV processes mediated by heavy Majorana neutrinos.
Fig.~\ref{fig:LNV} shows the predicted upper limits on the decay channels considered in our analysis, along with the corresponding experimental upper limits.

Among the decay modes, $B_c^- \to \pi^+ \mu^- \mu^-$ exhibits the largest branching ratio, $\mathcal{O}(10^{-8})$, which remains nearly unchanged for $M_N = 2$ and $3~\mathrm{GeV}$, while $B^- \to \pi^+ \mu^- \mu^-$ stays at $\mathcal{O}(10^{-11})$ for the same masses, indicating weak sensitivity to the heavy neutrino mass in this range. The decay $B_c^- \to J/\psi\pi^+ \mu^- \mu^-$ shows a strong suppression with increasing $M_N$, decreasing from $\mathcal{O}(10^{-10})$ to $\mathcal{O}(10^{-13})$. 
Overall, the results highlight a strong channel dependence, with $B_c$ modes showing higher sensitivity to heavy neutrino effects, while the mass dependence is governed by phase-space and hadronic structure effects.








\section{Discussions and Conclusion}
\label{Conclu_1}
In this work, we have explored the sensitivity of lepton-number violating $B$-meson decays in the presence of a heavy neutral lepton. We have first analyzed the purely leptonic decay $B^+ \to \ell^+ \nu$, where the final-state neutrino is identified as missing energy. In this framework, the Standard Model contribution interferes with possible effects from an on-shell heavy neutrino $N$, depending on its mass and mixing with active neutrinos.
We have determined the allowed parameter space in the $(|U_{\ell N}|^2, M_N)$ plane under both unitary and non-unitary mixing assumptions, with further separation into universal and flavor-dependent mixing scenarios. We have imposed all relevant constraints from leptonic $B$ decays, direct searches, and precision observables, retaining only regions consistent with current bounds.\\

Using these allowed regions, we have studied the $\Delta L = 2$ three- and four-body $B$-meson decays mediated by on-shell Majorana neutrinos, focusing on dimuon channels such as $B^- \to \pi^+ \mu^- \mu^-$, $B_c^- \to \pi^+ \mu^- \mu^-$, and $B_c^- \to J/\psi\, \pi^+ \mu^- \mu^-$. This approach has established a direct connection between constraints from leptonic two-body decays and the sensitivity of multi-body LNV channels.\\

Our results have shown that the predicted branching ratios are strongly suppressed, lying in the range $\mathcal{O}(10^{-13})$--$\mathcal{O}(10^{-8})$, with a clear dependence on the decay channel. In particular, $B_c$ modes have exhibited enhanced sensitivity to heavy neutrino effects compared to $B$ modes, highlighting their importance in probing the parameter space. Overall, rare $B$-meson decays have provided a complementary probe of heavy neutrino physics and have offered promising avenues for testing lepton number violation in future experiments.
  \section{Acknowledgments}
DP would like to acknowledge the support of the Prime Minister's Research Fellowship, provided by the Government of India.

\appendix
\section{ Angular conventions and Kinematics}
We consider only tree-level diagrams in the calculation of the RH neutrino total decay width. In the relevant mass range~$0.140\,\text{GeV}\leq M_{N}\leq 6\,\text{GeV}$, the following channels contribute to
the total decay width of the heavy neutrino:
\begin{itemize}
 \item $N$ $\to\ell^{-}P^{+}$, where $\ell=e,\mu,\tau$, and $P^{+}$ $=\pi^{+},\,K^{+},\,D^{+},\,D_{s}^{+},\,B^{+}$~(for $\ell=e,\mu$). 
\item $N\rightarrow\nu_{\ell}P^{0}$, where $\nu_{\ell}$ are the flavor eigenstates $\nu_{e},\,\nu_{\mu},\,\nu_{\tau}$ and $P^{0}=\pi^{0},\,\eta,\,\eta',\,\eta_{c}$.
\item $N\rightarrow\ell^{-}V^{+}$, where $\ell~=~e,\,\mu,\,\tau$, and $V^{+}=\rho^{+},
K^{*+},\,D^{*+},\,D_{s}^{*+},\,B^{*+}$~(for $\ell=e,\mu$).
\item $N\rightarrow\nu_{\ell}V^{0}$, where $\nu_{\ell}=\nu_{e},\,\nu_{\mu},\,\nu_{\tau}$ and $V^{0}=\rho^{0},\,\omega,\,\phi,\,J/\psi$. 
\item $N\rightarrow\ell_{1}^{-}\ell_{2}^{+}\nu_{\ell_{2}}$, where $\ell_{1},\,\ell_{2}=e,\,\mu,\,\tau$, $\ell_{1}\neq
\ell_{2}.$
\item $N\rightarrow\nu_{\ell_{1}}\ell_{2}^{-}\ell_{2}^{+}$, where $\ell_{1},\,\ell_{2}=e,\,\mu,\,\tau$. 
\item $N\rightarrow\nu_{\ell_{1}}\nu\overline{\nu}$, where $\nu_{\ell_{1}}=\nu_{e},\,\nu_{\mu},\,\nu_{\tau}$.
\end{itemize}
Hence, the total decay width is given by
\begin{equation}
\begin{split}
&\Gamma_{N}=\sum_{\ell,P^{+}}2\Gamma^{\ell P^{+}}+\sum_{\ell,P^{0}}\Gamma^{\nu_{\ell}P^{0}}+\sum_{\ell,V^{+}}2\Gamma^{\ell V^{+}}+\sum_{\ell,V^{0}}\Gamma^{\nu_{\ell}V^{0}}\\
&+\sum_{\ell_{1},\ell_{2}(\ell_{1}\neq\ell_{2})}2\Gamma^{\ell_{1}\ell_{2}\nu_{\ell_{2}}}+\sum_{\ell_{1},\ell_{2}}\Gamma^{\nu_{\ell_{1}}\ell_{2}\ell_{2}}+\sum_{\nu_{\ell_{1}}}\Gamma^{\nu_{\ell_{1}}\nu\overline{\nu}}.
\end{split}
\label{Total decay width of N}
\end{equation}
As the RH neutrino is Majorana, charge conjugate processes are also allowed, and the decay rate is the same; hence, the 2 factor is included for some of the channels.

\section{Semileptonic decay $B_c^- \to J/\psi \mu^-N$}  \label{appA}

The decay amplitude for the semileptonic subprocess $B_c^-(p_{B_c}) \to J/\psi(p_{J/\psi}) \mu^-(p_\mu)N(p_N)$  can be written as 
\begin{align}
{\cal M}(B_c^- \to J/\psi \mu^- N)
&= \frac{G_F}{\sqrt{2}}\,
V_{cb} U_{\mu N}
\langle J/\psi |\bar{b}\gamma^{\alpha}\gamma_5 c|B_c \rangle
\nonumber\\
& \times
\left[\bar{u}(p_\mu)\gamma_{\alpha}(1-\gamma_5)v(p_N)\right],
\label{AmpliA}
\end{align}

\noindent
where $G_F$ denotes  the Fermi constant, $V_{cb}$ represents the relevant CKM
matrix element, and $U_{\mu N}$ corresponds to the mixing parameter between the muon and heavy neutrino (sterile) $N$ in the charged current interaction
\cite{Atre:2009rg}. Within the framework of heavy quark spin symmetry, the hadronic current can
be expressed as \cite{Jenkins:1992nb,Colangelo:1999zn}
\begin{align}
\langle J/\psi |\bar{b}\gamma^{\alpha}\gamma_5 c|B_c \rangle
&= \sqrt{4 m_{B_c} m_{J/\psi}}\,
\Delta^{B_c \to J/\psi}\,
\epsilon^{\alpha *}.
\end{align}
\noindent Here transition is described by the single form factor $\Delta^{B_c \to J/\psi}$, where  $\epsilon^{\alpha}$ denotes the polarization vector of the $J/\psi$ meson. in  QCD relativistic potential
model framework \cite{Colangelo:1999zn}, this form factor is parametrized using a three-parameter expression given in Eq. \eqref{eq:FF_param}. 

The decay width can be expressed in terms of three-body phase space \cite{Olive_2014},
\begin{equation}
\Gamma(B_c^- \to J/\psi \mu^-N) = \dfrac{1}{32(2\pi)^{3} m_{B_c}^{3}}  \int_{t^{-}}^{t^{+}}  dt \int_{s^{-}}^{s^{+}} ds \ |\overline{\mathcal{M}}|^{2},
\end{equation}

\noindent where $|\overline{\mathcal{M}}|^{2}$ represents  the spin-averaged squared amplitude. The kinematic variables are defined as $s=(p_{J/\psi} + p_N)^{2}$ and $t=(p_{B_c} - p_{J/\psi})^2$. In the limit where charged lepton masses are neglected, the integration limits are given by $t^{-} = M_{N}^{2}$, $t^{+} = (m_{B_c} - m_{J/\psi})^{2}$ and
\begin{eqnarray}
s^{\pm}(t) =&& m_{B_c}^2 + M_N^2 - \dfrac{1}{2 t} \Big[ (t + m_{B_c}^2 - m_{J/\psi}^2)(t + M_N^2) \nonumber\\
&& \mp \ \lambda_t^{1/2} (M_N^2 - t) \ \Big],
\end{eqnarray}

\noindent where $\lambda_t \equiv \lambda(t,m_{B_c}^2,m_{J/\psi}^2)$,  with $\lambda(x,y,z)=x^{2}+y^{2}+z^{2}-2xy-2xz-2yz$ the usual kinematic K\"{a}llen function. After integration over the kinematical variable $s$, the decay width can be written as
\begin{align}
\Gamma(B_c^- \to J/\psi \mu^- N)
&= \frac{G_F^2}{32(2\pi)^3 m_{B_c}^3}
|V_{cb}|^2 |V_{\mu N}|^2  \nonumber\\
&\times
\int_{M_N^2}^{(m_{B_c}-m_{J/\psi})^2}
dt \; \mathcal{A}(t) ,
\label{BctopimuN}
\end{align}

\noindent with 
\begin{align}
\mathcal{A}(t)
&= \frac{8 m_{B_c}}{3 m_{J/\psi} t^3}
(\Delta^{B_c \to J/\psi})^2
(M_N^2 - t)^2 \lambda_t^{1/2} \nonumber \\
&\quad \times \Big[
\lambda_t (2 M_N^2 + t)
+ 6 m_{J/\psi}^2 t (M_N^2 + 2t)
\Big].
\label{At}
\end{align}

\noindent being a function of the squared transfer momentum $t$. The hadronic $B_c \to J/\psi$ transition is described in terms of a single form factor, $\Delta^{B_c \to J/\psi}$~\cite{Jenkins:1992nb,Colangelo:1999zn}, which, within the QCD relativistic potential model, is parametrized as
\begin{equation}
\Delta^{B_c \to J/\psi}(y) 
= \Delta(0)\left[1 - \rho^2 (y-1) + c (y-1)^2 \right],
\label{eq:FF_param}
\end{equation}
with $\Delta(0)=0.94$, $\rho^2 = 2.9$, and $c=3$~\cite{Colangelo:1999zn}. The recoil variable $y$ is related to the momentum transfer by
\begin{equation}
y=\frac{m_{B_c}^2 + m_{J/\psi}^2 - t}{2\,m_{B_c} m_{J/\psi}}\,.
\end{equation}

\bibliographystyle{ieeetr}
\bibliography{LNV}
\end{document}